\pdfoutput=1
\RequirePackage{ifpdf}
\ifpdf 
\documentclass[pdftex]{sigma}
\else
\documentclass{sigma}
\fi

\numberwithin{equation}{section}

\begin{document}


\newcommand{\arXivNumber}{1704.04078}

\renewcommand{\thefootnote}{}

\renewcommand{\PaperNumber}{058}

\FirstPageHeading

\ShortArticleName{Relativistic DNLS and Kaup--Newell Hierarchy}

\ArticleName{Relativistic DNLS and Kaup--Newell Hierarchy\footnote{This paper is a~contribution to the Special Issue on Symmetries and Integrability of Dif\/ference Equations. The full collection is available at \href{http://www.emis.de/journals/SIGMA/SIDE12.html}{http://www.emis.de/journals/SIGMA/SIDE12.html}}}

\Author{Oktay K.~PASHAEV~$^\dag$ and Jyh-Hao LEE~$^\ddag$}

\AuthorNameForHeading{O.K.~Pashaev and J.-H.~Lee}

\Address{$^\dag$~Department of Mathematics, Izmir Institute of Technology, Urla-Izmir 35430, Turkey}
\EmailD{\href{mailto:oktaypashaev@iyte.edu.tr}{oktaypashaev@iyte.edu.tr}}
\URLaddressD{\url{http://www.iyte.edu.tr/~oktaypashaev/}}

\Address{$^\ddag$~Institute of Mathematics, Academia Sinica, Taipei 10617, Taiwan}
\EmailD{\href{mailto:leejh@math.sinica.edu.tw}{leejh@math.sinica.edu.tw}}

\ArticleDates{Received April 14, 2017, in f\/inal form July 18, 2017; Published online July 25, 2017}

\Abstract{By the recursion operator of the Kaup--Newell hierarchy we construct the rela\-ti\-vistic derivative NLS (RDNLS) equation and the corresponding Lax pair. In the nonrelativistic limit $c \rightarrow \infty$ it reduces to DNLS equation and preserves integrability at any order of relativistic corrections. The compact explicit representation of the linear problem for this equation becomes possible due to notions of the $q$-calculus with two bases, one of which is the recursion operator, and another one is the spectral parameter.}

\Keywords{Kaup--Newell hierarchy; relativistic DNLS; $q$-calculus; recursion operator}

\Classification{35Q55; 37K10}

\renewcommand{\thefootnote}{\arabic{footnote}}
\setcounter{footnote}{0}

\section{Introduction}

The Derivative NLS (DNLS) equation was introduced in plasma physics as descriptive of weakly nonlinear and dispersive parallel MHD waves \cite{Mio,M}. The equation is integrable and belongs to the Kaup--Newell (KN) hierarchy \cite{KN, LEE}. As a model of one-dimensional anyons it was studied in \cite{Aglietti,Jackiw,Min}. Solving problem of chiral soliton in quantum potential, the ${\rm SL}(2,{\mathbb R})$ version of this equation was derived from the Chern--Simons gauge theory in $2+1$ dimensions~\cite{chiral}. The ${\rm SL}(2,{\mathbb R})$ version of DNLS, also known as DRD or resonant DNLS, has solutions in the form of soliton resonances with chiral properties \cite{LLP,ANZIAM}. The ${\rm SL}(2,{\mathbb R})$ KN hierarchy, reformulated in~\cite{Yan} by modif\/ied spectral problem, was applied in~\cite{FLP, LeePash} to obtain resonance solitons for modif\/ied KP equation.

In the present paper we are going to apply DNLS hierarchy to construct the relativistic DNLS equation as an integrable nonlinearization of the semirelativistic Schr\"odinger equation. The relativistic version of the NLS equation, based on the Zakharov--Shabat hierarchy was derived before in~\cite{Pashaev}. We show that the RDNLS equation in the nonrelativistic limit reduces to the DNLS equation, and at any order of relativistic corrections~$1/c^2$ it produces an integrable model. To represent the linear problem in a compact explicit form we use notions of $q$-number and $q$-derivative from the $q$-calculus with two bases, one of which is the recursion operator, and another one is the spectral parameter.

The paper is organized as follows. In Section~\ref{section2} we review ${\rm SL}(2,{\mathbb R})$ KN hierarchy in formulation of~\cite{Yan}. This formulation is linear in the spectral parameter, in contrast to the original KN paper with the quadratic dependence. We found the f\/irst one very convenient for our calculations, and problem of equivalency with the second one still requires to be clarif\/ied. We represent the corresponding linear problem in terms of $q$-calculus with two bases: one of which is the recursion operator and another one is the spectral parameter. For this hierarchy in Section~\ref{section3} we derive equation and the linear problem with an arbitrary dispersion. Section~\ref{section4} is devoted to the DNLS hierarchy. In Section~\ref{section4.1} we study DNLS hiearchy as integrable deformations of the linear Schr\"odinger equation and the corresponding linear hierarchy of higher-order equations. Arbitrary dispersive DNLS and corresponding linear problem are the subject of Section~\ref{section4.2}. In Section~\ref{section5} we introduce the relativistic DNLS, its nonrelativistic reductions and integrable corrections at any order. Implications for the soliton solutions and the relativistic character of time dependence are discussed brief\/ly in conclusions.

\section[${\rm SL}(2,{\mathbb R})$ KN hierarchy]{$\boldsymbol{{\rm SL}(2,{\mathbb R})}$ KN hierarchy}\label{section2}

 Here we brief\/ly review the KN hierarchy in the form of \cite{Yan}, with the linear dependence on the spectral parameter for~$U$. The linear problem for the zero curvature equation
 \begin{gather*}
 U_t -V_x + [U, V] = 0
 \end{gather*}
 is determined by the system of linear equations
 \begin{gather*}
 \phi_x = U \phi, \qquad \phi_t = V \phi,
\end{gather*}
where
\begin{gather*}
U(x,t;\lambda) = \left( \begin{matrix} \lambda & q(x,t) \\ \lambda r(x,t) & -\lambda \end{matrix} \right),
\end{gather*}
and
\begin{gather*}
V(x,t;\lambda ) = \left( \begin{matrix} A & B \\ C & -A \end{matrix} \right).
\end{gather*}
It gives us the system of equations
\begin{gather}
q_t = B_x - 2\lambda B + 2 q A ,\label{q}\\
\lambda r_t = C_x + 2\lambda C - 2 \lambda r A ,\label{r}\\
A_x = q C - \lambda r B.\nonumber
\end{gather}
By expanding in spectral parameter
\begin{gather*}
A = \sum^{N}_{m=0} a_m \lambda^{N+1-m} = a_0 \lambda^{N+1} + a_1 \lambda^{N} + a_2 \lambda^{N-1}+\dots + a_N \lambda,\\
B = \sum^{N+1}_{m=0} b_m \lambda^{N+1-m} = b_0 \lambda^{N+1} + b_1 \lambda^{N} + b_2 \lambda^{N-1}+\dots + b_N \lambda + b_{N+1},\\
C = \sum^{N}_{m=0} c_m \lambda^{N+1-m} = c_0 \lambda^{N+1} + c_1 \lambda^{N} + c_2 \lambda^{N-1}+\dots + c_N \lambda,
\end{gather*}
from the last equations we get recursions
\begin{gather}
\frac{d}{dx}b_{k+1} - 2 b_{k+2} + 2 q a_{k+1} = 0, \qquad k = 0,\dots ,N-1 ,\label{bk}\\
\frac{d}{dx}c_{k} + 2 c_{k+1} - 2 r a_{k+1} = 0, \qquad k = 0,\dots ,N-1 ,\label{ck}\\
\frac{d}{dx}a_{k} = q c_{k} - r b_{k+1}, \qquad k = 0,\dots ,N, \label{ak}\\
\frac{d}{dx}b_{0} - 2 b_{1} + 2 q a_{0} = 0,\nonumber
\end{gather}
$b_0 = 0$, $a_0 = 1$, $c_0 = r$. Then, as follows $ b_1 = q$. From (\ref{ak}) we have
\begin{gather}
a_k = \big(\partial^{-1} q, - \partial^{-1} r\big) \left( \begin{matrix} c_k \\ b_{k+1} \end{matrix}\right) \label{akGk}
\end{gather}
and combining (\ref{bk}) and (\ref{ck})
\begin{gather*}
\frac{d}{dx} \left( \begin{matrix} c_k \\ b_{k+1} \end{matrix}\right) + 2 \left( \begin{matrix} c_{k+1} \\ -b_{k+2} \end{matrix}\right)
-2 a_{k+1}\left( \begin{matrix} r \\ -q \end{matrix}\right) = 0.
\end{gather*}
In terms of
\begin{gather}
G_k = \left( \begin{matrix} c_k \\ b_{k+1} \end{matrix}\right)\label{Gk}
\end{gather}
we get
\begin{gather*}
\left( \begin{matrix} 1 - r \partial^{-1} q& r \partial^{-1} r\\ q \partial^{-1}q & -1 - q \partial^{-1} r\end{matrix}\right) G_{k+1} =
 - \frac{1}{2} \left( \begin{matrix} \partial & 0\\ 0 & \partial \end{matrix}\right) G_k,
\end{gather*}
$k = 0, \dots , N-1$. This relation can be rewritten as
\begin{gather}
G_{k+1} = L G_k = L^2 G_{k-1} = \dots = L^{k+1} G_0,\qquad G_0 = \left( \begin{matrix} r \\ q\end{matrix}\right) ,\label{recurence}
\end{gather}
where the recursion operator is
\begin{gather}
L = \frac{1}{2}\left( \begin{matrix} -\partial - r \partial^{-1} q \partial & -r \partial^{-1} r\partial\\
-q \partial^{-1}q \partial & \partial - q \partial^{-1} r \partial\end{matrix}\right) .\label{L}
\end{gather}
Substituting to (\ref{q}) and (\ref{r}) we get the $N$-th f\/low of ${\rm SL}(2,{\mathbb R})$ KN hierarchy \cite{Yan}
\begin{gather}
\left( \begin{matrix} q \\ r\end{matrix}\right)_{t_N} = J L^N \left( \begin{matrix} r \\ q\end{matrix}\right), \label{knh}
\end{gather}
where
\begin{gather*}
J = \left( \begin{matrix} 0 & \partial \\ \partial & 0\end{matrix}\right).
\end{gather*}

As was shown in \cite{Yan}, this set of equations possesses inf\/initely many commuting symmetries associated with every f\/low~$t_N$. This allows us to combine these f\/lows in an arbitrary linear combination form to construct new equations def\/ined on this hierarchy.

Unfortunately in \cite{Yan} no explicit form of the Lax pair in terms of recursion operator is given. Up to our knowledge it is also not known
before for the AKNS hierarchy, and for that case it was f\/irst derived in~\cite{Pashaev}. Here we are going to complete this part and give simple and compact form of the Lax pair for the KN hierarchy. For this reason we need to introduce some notions from $q$-calculus,
which allows us to get explicit formula for the Lax pair. For NLS hierarchy it was constructed before in~\cite{Pashaev}.

The nonsymmetric $q$-number with one base is def\/ined by
\begin{gather*}
[n]_q = \frac{q^n - 1}{q-1}
\end{gather*}
and it is a particular case of the $pq$-number with two bases
\begin{gather*}
[n]_{pq} = \frac{p^n - q^n}{p-q}.
\end{gather*}
We can extend these def\/initions to operator $q$-numbers with a base as an operator. In particular we will use next notations, for the nonsymmetric operator $q$-number
\begin{gather}
[n]_Q = \frac{1 - Q^n}{1 - Q} \equiv I + Q + Q^2 + \dots + Q^{n-1} \label{nA}
\end{gather}
and for the two bases operator $q$-number
\begin{gather*}
[n]_{PQ} = \frac{P^n - Q^n}{P-Q} \equiv P^{n-1} Q + P^{n-2} Q^{2} + \dots + P^2 Q^{n-2} + P Q^{n-1}. 
\end{gather*}

For the Lax pair members we have
\begin{gather*}
\left( \begin{matrix} B \\ C\end{matrix}\right) = \left( \begin{matrix} 0 & 1 \\ 1 & 0\end{matrix}\right)\left( \begin{matrix} C \\ B\end{matrix}\right)
= \sigma_1 \sum^{N+1}_{m=0} \left( \begin{matrix} c_m \\ b_m\end{matrix}\right) \lambda^{N+1-m},
\end{gather*}
where $c_{N+1} = 0$ and $b_0 =0$. Explicitly
 \begin{gather*}
\left( \begin{matrix} C \\ B\end{matrix}\right) = \sum^{N+1}_{m=0} \left( \begin{matrix} c_m \\ b_m\end{matrix}\right) \lambda^{N+1-m} =
\left( \begin{matrix} c_0 \lambda^{N+1} + c_1 \lambda^N + \dots + c_N \lambda \\ b_1 \lambda^N + \dots + b_N \lambda + b_{N+1} \end{matrix}\right)
\end{gather*}
or by combining terms as
 \begin{gather*}
\left( \begin{matrix} C \\ B\end{matrix}\right) = \left( \begin{matrix} c_0 \lambda^{N+1} \\ b_1 \lambda^N\end{matrix}\right) + \left( \begin{matrix} c_1 \lambda^{N} \\ b_2 \lambda^{N-1}\end{matrix}\right) + \dots +
 \left( \begin{matrix} c_N \lambda \\ b_{N+1}\end{matrix}\right)
\end{gather*}
we have
 \begin{gather*}
\left( \begin{matrix} C \\ B\end{matrix}\right) = \left( \begin{matrix} c_0 \lambda \\ b_1 \end{matrix}\right) \lambda^N + \left( \begin{matrix} c_1 \lambda \\ b_2 \end{matrix}\right) \lambda^{N-1} + \dots +
 \left( \begin{matrix} c_N \lambda \\ b_{N+1}\end{matrix}\right) \\
 \hphantom{\left( \begin{matrix} C \\ B\end{matrix}\right)}{}
 = \left( \begin{matrix} \lambda & 0 \\ 0 & 1\end{matrix}\right) \left( \left( \begin{matrix} c_0 \\ b_1 \end{matrix}\right) \lambda^N + \left( \begin{matrix} c_1 \\ b_2 \end{matrix}\right) \lambda^{N-1} + \dots +
 \left( \begin{matrix} c_N \\ b_{N+1}\end{matrix}\right)\right).
\end{gather*}
In terms of (\ref{Gk}) then we get
\begin{gather*}
\left( \begin{matrix} C \\ B\end{matrix}\right) = \left( \begin{matrix} \lambda & 0 \\ 0 & 1\end{matrix}\right) \big(\lambda^N G_0 + \lambda^{N-1} G_1 + \dots +\lambda G_{N-1} + G_N\big)
\end{gather*}
or by using (\ref{recurence}) and recursion operator (\ref{L})
\begin{gather*}
\left( \begin{matrix} C \\ B\end{matrix}\right) = \left( \begin{matrix} \lambda & 0 \\ 0 & 1\end{matrix}\right) \big(\lambda^N + \lambda^{N-1} L +
\dots +\lambda L^{N-1} + L^N\big) G_0
\\
\hphantom{\left( \begin{matrix} C \\ B\end{matrix}\right)}{} = \left( \begin{matrix} \lambda^{N+1} & 0 \\ 0 & \lambda^N\end{matrix}\right) \left(1 + \frac{L}{\lambda} + \frac{L^2}{\lambda^2}+
\dots + \frac{L^N}{\lambda^N}\right) G_0.
\end{gather*}
Finally, due to (\ref{nA}) we can rewrite the result in a short form as an operator $q$-number or as a~$q$-derivative, with the scaled recursion operator $L/\lambda$ as a base,
\begin{gather*}
\left( \begin{matrix} C \\ B\end{matrix}\right) = \left( \begin{matrix} \lambda^{N+1} & 0 \\ 0 & \lambda^N\end{matrix}\right) [N+1]_{L/\lambda}\left( \begin{matrix} r \\ q\end{matrix}\right) = \left( \begin{matrix} \lambda & 0 \\ 0 & 1\end{matrix}\right) D_{L/\lambda} \lambda^{N+1}\left( \begin{matrix} r \\ q\end{matrix}\right),
\end{gather*}
where the operator $q$-derivative is def\/ined as
\begin{gather}
D_{L/\lambda} f(\lambda) = \frac{f(L) - f(\lambda)}{L - \lambda}.\label{D}
\end{gather}
In explicit form we have
\begin{gather*}
\left( \begin{matrix} C \\ B\end{matrix}\right) = \left( \begin{matrix} \lambda & 0 \\ 0 & 1\end{matrix}\right)
\frac{L^{N+1} - \lambda^{N+1}}{L - \lambda}\left( \begin{matrix} r \\ q\end{matrix}\right).
\end{gather*}

In a similar way, due to (\ref{akGk}) we obtain
\begin{gather*}
A = \sum^N_{m=0} a_m \lambda^{N+1-m} = \lambda^{N+1} + \sum^N_{m=1} \lambda^{N+1-m} \big(\partial^{-1} q, - \partial^{-1} r\big) G_m \\
 \hphantom{A}{} = \lambda^{N+1} + \big(\partial^{-1} q, - \partial^{-1} r\big) \sum^N_{m=1} \big(\lambda^N L + \lambda^{N-1} L^2 + \dots + \lambda L^N\big) G_0.
\end{gather*}
It can be rewritten in terms of the operator $q$-number and the $q$-derivative as follows
\begin{gather*}
A = \lambda^{N+1} + \big(\partial^{-1} q, - \partial^{-1} r\big) \lambda^N L [N]_{L/\lambda}\left( \begin{matrix} r \\ q\end{matrix}\right)\\
\hphantom{A}{} = \lambda^{N+1} + \big(\partial^{-1} q, - \partial^{-1} r\big) \lambda L D_{L/\lambda} \lambda^N \left( \begin{matrix} r \\ q\end{matrix}\right)
\end{gather*}
or
\begin{gather*}
A = \lambda^{N+1} + \big(\partial^{-1} q, - \partial^{-1} r\big) \lambda L \frac{L^N - \lambda^N}{L - \lambda} \left( \begin{matrix} r \\ q\end{matrix}\right).
\end{gather*}

\section{Arbitrary dispersion and KN hierarchy}\label{section3}

As we have seen in previous section, the $N$-th f\/low of ${\rm SL}(2,{\mathbb R})$ KN hierarchy is described by equation
 \begin{gather*}
\left( \begin{matrix} q \\ r\end{matrix}\right)_{t_N} = J L^N \left( \begin{matrix} r \\ q\end{matrix}\right),
\end{gather*}
which is equivalent to the zero-curvature condition
\begin{gather}
\partial_{t_{N}} U - \partial_x V_N + [U, V_N] = 0, \label{zcN}
\end{gather}
where
\begin{gather*}
U(x,t;\lambda) = \left( \begin{matrix} \lambda & q(x,t) \\ \lambda r(x,t) & -\lambda \end{matrix} \right),\qquad
V_N(x,t;\lambda) = \left( \begin{matrix} A_N & B_N \\ C_N & -A_N \end{matrix} \right),
\\
\left( \begin{matrix} C_N \\ B_N\end{matrix}\right) = \left( \begin{matrix} \lambda & 0 \\ 0 & 1\end{matrix}\right)
\frac{L^{N+1} - \lambda^{N+1}}{L - \lambda}\left( \begin{matrix} r \\ q\end{matrix}\right),
\\
A_N = \lambda^{N+1} + \big(\partial^{-1} q, - \partial^{-1} r\big) \lambda L \frac{L^N - \lambda^N}{L - \lambda} \left( \begin{matrix} r \\ q\end{matrix}\right).
\end{gather*}

Now, let us introduce a new time variable $t$, determined by this hierarchy as
\begin{gather*}
\frac{\partial}{\partial t} = \sum^\infty_{N = 0}\nu_N \frac{\partial}{\partial t_N}.
\end{gather*}
Then, equations of motion are
\begin{gather}
\left( \begin{matrix} q \\ r\end{matrix}\right)_{t}
=\sum^\infty_{N = 0}\nu_N \left( \begin{matrix} q \\ r\end{matrix}\right)_{t_N} = J\sum^\infty_{N = 0}\nu_N L^N \left( \begin{matrix} r \\ q\end{matrix}\right) \label{qr}
\end{gather}
or
\begin{gather*}
\left( \begin{matrix} q \\ r\end{matrix}\right)_{t} = J F(L) \left( \begin{matrix} r \\ q\end{matrix}\right),
\end{gather*}
where function
\begin{gather*}
F(z) = \sum^\infty_{N = 0}\nu_N z^N
\end{gather*}
is the symbol of operator $F(L)$. The linear problem and the zero-curvature condition corresponding to (\ref{qr}) are determined by the related sum of equations~(\ref{zcN})
\begin{gather*}
\partial_{t} U - \partial_x V + [U, V] = 0, 
\end{gather*}
where
\begin{gather*}
V = \left( \begin{matrix} A & B \\ C & -A \end{matrix} \right) = \sum^\infty_{N = 0}\nu_N V_N.
\end{gather*}
Then
\begin{gather*}
\left( \begin{matrix} C \\ B\end{matrix}\right) = \left( \begin{matrix} \lambda & 0 \\ 0 & 1\end{matrix}\right)
\frac{L \sum\limits^\infty_{N = 0} \nu_N L^{N} - \lambda \sum\limits^\infty_{N = 0} \nu_N \lambda^{N}}{L - \lambda}\left( \begin{matrix} r \\ q\end{matrix}\right),
\end{gather*}
or in compact form
\begin{gather*}
\left( \begin{matrix} C \\ B\end{matrix}\right) = \left( \begin{matrix} \lambda & 0 \\ 0 & 1\end{matrix}\right)
\frac{L F(L) - \lambda F(\lambda)}{L - \lambda}\left( \begin{matrix} r \\ q\end{matrix}\right).
\end{gather*}
For
\begin{gather*}
A = \sum^\infty_{N = 0} \nu_N A_N = \lambda \sum^\infty_{N = 0} \nu_N \lambda^{N} + \big(\partial^{-1} q, - \partial^{-1} r\big) \lambda L \sum^\infty_{N = 0} \nu_N\frac{L^N - \lambda^N}{L - \lambda} \left( \begin{matrix} r \\ q\end{matrix}\right)
\end{gather*}
we f\/ind expression as
\begin{gather*}
A = \lambda F(\lambda) + \big(\partial^{-1} q, - \partial^{-1} r\big) \lambda L \frac{F(L) - F(\lambda)}{L - \lambda} \left( \begin{matrix} r \\ q\end{matrix}\right).
\end{gather*}
In terms of the operator $q$-derivative (\ref{D}), the connection components become
\begin{gather*}
\left( \begin{matrix} C \\ B\end{matrix}\right) = \left( \begin{matrix} \lambda & 0 \\ 0 & 1\end{matrix}\right)
 D_{L/\lambda} \lambda F(\lambda)\left( \begin{matrix} r \\ q\end{matrix}\right),\\
A = \lambda F(\lambda) + \big(\partial^{-1} q, - \partial^{-1} r\big) \lambda L D_{L/\lambda} F(\lambda) \left( \begin{matrix} r \\ q\end{matrix}\right).
\end{gather*}

\subsection{RD system}\label{section3.1}
As an example we consider the linear function case $F(x) = x$ and corresponding equations of motion
\begin{gather*}
\left( \begin{matrix} q \\ r\end{matrix}\right)_{t} = J F(L) \left( \begin{matrix} r \\ q\end{matrix}\right) = J L \left( \begin{matrix} r \\ q\end{matrix}\right).
\end{gather*}
This gives the derivative reaction-dif\/fusion system (DRD)
\begin{gather*}
q_t = \frac{1}{2}\big(q_{xx} - \big(q^2 r\big)_x\big), \\ 
-r_t = \frac{1}{2}\big(r_{xx} + \big(r^2 q\big)_x\big), 
\end{gather*}
with the linear problem
\begin{gather*}
\left( \begin{matrix} C \\ B\end{matrix}\right) = \left( \begin{matrix} \lambda & 0 \\ 0 & 1\end{matrix}\right)
\frac{L^2 - \lambda^2}{L - \lambda}\left( \begin{matrix} r \\ q\end{matrix}\right) = \left( \begin{matrix} \lambda & 0 \\ 0 & 1\end{matrix}\right)
(L + \lambda)\left( \begin{matrix} r \\ q\end{matrix}\right),
\end{gather*}
or
\begin{gather*}
\left( \begin{matrix} C \\ B\end{matrix}\right) =
 \left( \begin{matrix} \lambda^2 r -\frac{1}{2}\lambda r_x - \frac{1}{2}\lambda r^2 q \\ \lambda q +\frac{1}{2} q_x - \frac{1}{2} q^2 r \end{matrix}\right).
\end{gather*}
Then for $A$ we have
\begin{gather*}
A = \lambda^2 + \big(\partial^{-1} q, - \partial^{-1} r\big) \lambda L \left( \begin{matrix} r \\ q\end{matrix}\right),
\end{gather*}
and thus
\begin{gather*}
A = \lambda^2 - \frac{1}{2}\lambda r q.
\end{gather*}
As a result, we get the following zero curvature potentials
\begin{gather*}
U(x,t;\lambda) = \left( \begin{matrix} \lambda & q \\ \lambda r & -\lambda \end{matrix} \right),\\
V(x,t;\lambda) = \left( \begin{matrix} \lambda^2 - \frac{1}{2}\lambda r q & \lambda q +\frac{1}{2} q_x - \frac{1}{2} q^2 r \\ \lambda^2 r -\frac{1}{2}\lambda r_x - \frac{1}{2}\lambda r^2 q & -\lambda^2 + \frac{1}{2}\lambda r q \end{matrix} \right).
\end{gather*}
This DRD system was studied in \cite{LLP,ANZIAM} by Hirota's bilinear method and the soliton solutions with resonant interaction were derived. Combined with the next hierarchy f\/low, it produces the MKP-II equation, for which the resonant solitons was obtained and studied as well in~\cite{FLP, LeePash}.
\section{DNLS hierarchy}\label{section4}

The DNLS hierarchy can be derived from ${\rm SL}(2,{\mathbb R})$ KN hierarchy (\ref{knh}) by formal substitutions. As a f\/irst step we replace operator $L$ by
\begin{gather*}
{\cal L} = \sigma_1 L \sigma_1,
\end{gather*}
so that the hierarchy (\ref{knh}) becomes
\begin{gather*}
\left( \begin{matrix} q \\ r\end{matrix}\right)_{t_N} = \partial {\cal L}^N \left( \begin{matrix} q \\ r\end{matrix}\right).
\end{gather*}
Then, we substitute
\begin{gather*}
\frac{\partial}{\partial t_N} \rightarrow i\frac{\partial}{\partial t_N}, \qquad
\frac{\partial}{\partial x} \rightarrow -i\frac{\partial}{\partial x}, \qquad q \rightarrow \psi,\qquad r \rightarrow \kappa^2 \bar \psi,
\end{gather*}
where $\kappa^2 = \pm 1$. So, the hierarchy can be rewritten as
\begin{gather*}
i \frac{\partial}{\partial t_N}\left( \begin{matrix} \psi \\ \kappa^2\bar\psi\end{matrix}\right) = -i \frac{\partial}{\partial x}
 {\cal L}^N \left( \begin{matrix} \psi \\ \kappa^2\bar\psi \end{matrix}\right),
\end{gather*}
where
\begin{gather*}
{\cal L} = \frac{1}{2}\left( \begin{matrix} -i\partial - \kappa^2\psi \partial^{-1} \bar\psi \partial & -\psi \partial^{-1} \psi\partial\\
-\bar\psi \partial^{-1}\bar\psi \partial & i\partial - \kappa^2 \bar\psi \partial^{-1} \psi \partial\end{matrix}\right)
\end{gather*}
or
\begin{gather*}
i\sigma_3 \frac{\partial}{\partial t_N}\left( \begin{matrix} \psi \\ \bar\psi\end{matrix}\right) = -i \sigma_3\frac{\partial}{\partial x}
 \left( \begin{matrix} 1&0 \\0 &\kappa^2 \end{matrix}\right){\cal L}^N \left( \begin{matrix} 1&0 \\0 &\kappa^2 \end{matrix}\right) \left( \begin{matrix} \psi \\ \bar\psi \end{matrix}\right).
\end{gather*}
By introducing
\begin{gather*}
{\cal M}\equiv \left( \begin{matrix} 1&0 \\0 &\kappa^2 \end{matrix}\right){\cal L} \left( \begin{matrix} 1&0 \\0 &\kappa^2 \end{matrix}\right),
\end{gather*}
f\/inally we get the DNLS hierarchy
\begin{gather}
i\sigma_3 \frac{\partial}{\partial t_N}\left( \begin{matrix} \psi \\ \bar\psi\end{matrix}\right) = -i \sigma_3\frac{\partial}{\partial x}
 {\cal M}^N \left( \begin{matrix} \psi \\ \bar\psi \end{matrix}\right),\label{DNLSH}
\end{gather}
with the recursion operator
\begin{gather}
{\cal M} = \frac{1}{2}\left( \begin{matrix} -i\partial - \kappa^2\psi \partial^{-1} \bar\psi \partial & -\kappa^2\psi \partial^{-1} \psi\partial\\
-\kappa^2\bar\psi \partial^{-1}\bar\psi \partial & i\partial - \kappa^2 \bar\psi \partial^{-1} \psi \partial\end{matrix}\right).\label{M}
\end{gather}

\subsection{Integrable deformation of linear Schr\"odinger equation}\label{section4.1}
In the linear limit, which can be formally taken as $\kappa^2 \rightarrow 0$, the recursion operator becomes just half of the momentum operator
\begin{gather}
 {\cal M}_0 = \frac{1}{2} \left(-i \sigma_3\frac{\partial}{\partial x}\right) \label{M0}
\end{gather}
and the f\/irst f\/low of hierarchy reduces to the linear Schr\"odinger equation and its complex conjugate, with $\hbar = 1$, $m =1$,
\begin{gather}
i\sigma_3 \frac{\partial}{\partial t_1}\left( \begin{matrix} \psi \\ \bar\psi\end{matrix}\right) = -\frac{1}{2}\frac{\partial^2}{\partial x^2}
 \left( \begin{matrix} \psi \\ \bar\psi \end{matrix}\right).\label{lS}
\end{gather}
Then, for the nonlinear case $\kappa^2 \neq 1$ we get the following f\/irst f\/low equation
\begin{gather*}
i\sigma_3 \frac{\partial}{\partial t_1}\left( \begin{matrix} \psi \\ \bar\psi\end{matrix}\right) = -\frac{1}{2}\frac{\partial^2}{\partial x^2}
 \left( \begin{matrix} \psi \\ \bar\psi \end{matrix}\right) + \frac{1}{2}\kappa^2 i\sigma_3 \frac{\partial}{\partial x} |\psi|^2
 \left( \begin{matrix} \psi \\ \bar\psi \end{matrix}\right) .
\end{gather*}
which is just the DNLS equation
\begin{gather*}
i \frac{\partial}{\partial {t_{1}}} \psi = - \frac{1}{2}\frac{\partial^2}{\partial x^2}\psi + i\frac{1}{2}\kappa^2 \frac{\partial}{\partial x}
\big(|\psi|^2 \psi \big).
\end{gather*}
The above consideration allows us to consider DNLS as the specif\/ic nonlinearization of the linear Schr\"odinger equation. Starting from the classical dispersion with Hamiltonian function
\begin{gather*}
E(p) = \frac{p^2}{2m},
\end{gather*}
by the f\/irst quantization rules $E \rightarrow i\hbar \partial/\partial t$, $p \rightarrow -i\hbar \partial/\partial x$, we get the linear Schr\"odinger equation
\begin{gather*}
i\hbar \frac{\partial}{\partial t}\psi = \frac{1}{2m}\left(-i\hbar \frac{\partial}{\partial x} \right)^2 \psi.
\end{gather*}
Combining this equation and its complex conjugate as in (\ref{lS}) ($\hbar = 1$, $m=1$), we can rewrite it in the form
\begin{gather*}
i\sigma_3 \frac{\partial}{\partial t}\left( \begin{matrix} \psi \\ \bar\psi\end{matrix}\right) = -i \sigma_3\frac{\partial}{\partial x}
 {\cal M}_0 \left( \begin{matrix} \psi \\ \bar\psi \end{matrix}\right).
\end{gather*}
Then, the DNLS model, as a nonlinearization of the linear Schr\"odinger equation, appears by replacement of the linear recursion operator
${\cal M}_0$~(\ref{M0}) by the nonlinear one, the operator~${\cal M}$ in~(\ref{M}), so that
\begin{gather*}
i\sigma_3 \frac{\partial}{\partial t}\left( \begin{matrix} \psi \\ \bar\psi\end{matrix}\right) = -i \sigma_3\frac{\partial}{\partial x}
 {\cal M} \left( \begin{matrix} \psi \\ \bar\psi \end{matrix}\right).
\end{gather*}
This procedure can be extended to the DNLS hierarchy. In the formal limit $\kappa^2 \rightarrow 0$ of DNLS hierarchy~(\ref{DNLSH}) we get the linear Schrodinger hierarchy
\begin{gather}
i\sigma_3 \frac{\partial}{\partial t_N}\left( \begin{matrix} \psi \\ \bar\psi\end{matrix}\right) = -i \sigma_3\frac{\partial}{\partial x}
 {{\cal M}_0}^N \left( \begin{matrix} \psi \\ \bar\psi \end{matrix}\right) = \frac{1}{2^N}\left(-i \sigma_3\frac{\partial}{\partial x}\right)^{N+1}
 \left( \begin{matrix} \psi \\ \bar\psi \end{matrix}\right) .\label{LSH}
\end{gather}
It corresponds to the f\/irst quantization of a classical system with dispersion
\begin{gather*}
E(p) = \frac{1}{2^N} p^{N+1}.
\end{gather*}
Then, following in the opposite direction, from this dispersion we get the linear Schr\"odinger hierarchy. By nonlinearization in (\ref{LSH}) we replace ${\cal M}_0$ by ${\cal M}$ and obtain the DNLS hierarchy:
 \begin{gather*}
i\sigma_3 \frac{\partial}{\partial t_N}\left( \begin{matrix} \psi \\ \bar\psi\end{matrix}\right) = -i \sigma_3\frac{\partial}{\partial x}
 {\cal M}^{-1}{\cal M}^{N+1} \left( \begin{matrix} \psi \\ \bar\psi \end{matrix}\right).
\end{gather*}
It is convenient to have it in the present form, ref\/lecting the same power for the classical momentum and the recursion operator.

\subsection{Arbitrary dispersive DNLS}\label{section4.2}
Now we are ready to consider a system with an arbitrary classical dispersion $E(p)$ and the Hamiltonian function
\begin{gather}
H(p) = E_0 + E_1 p + E_2 p^2 + \dots + E_N p^N + \cdots. \label{dispersion}
\end{gather}
In a more general case it is possible to have summation also in negative powers of~$p$ as a~Laurent series expansion, which will require to use the negative DNLS hierarchy f\/lows. The f\/irst quantized linear Schr\"odinger equation corresponding to this dispersion is
\begin{gather*}
i \frac{\partial}{\partial t} \psi = H\left(-i\frac{\partial}{\partial x}\right) \psi.
\end{gather*}
Combining it with its complex conjugate we have
 \begin{gather*}
i\sigma_3 \frac{\partial}{\partial t}\left( \begin{matrix} \psi \\ \bar\psi\end{matrix}\right) = \left(E_0 + E_1\left(-i \sigma_3\frac{\partial}{\partial x}\right) +\dots + E_N\left(-i \sigma_3\frac{\partial}{\partial x}\right)^N+\cdots \right) \left( \begin{matrix} \psi \\ \bar\psi \end{matrix}\right),
\end{gather*}
or in terms of operator ${\cal M}_0$~(\ref{M0})
 \begin{gather*}
i\sigma_3 \frac{\partial}{\partial t}\left( \begin{matrix} \psi \\ \bar\psi\end{matrix}\right) = -i\sigma_3 \frac{\partial}{\partial x}
\big(E_0 (2 {\cal M}_0)^{-1} + E_1 +\dots + E_{N+1}(2{\cal M}_0)^N+\cdots \big) \left( \begin{matrix} \psi \\ \bar\psi \end{matrix}\right)\\
\hphantom{i\sigma_3 \frac{\partial}{\partial t}\left( \begin{matrix} \psi \\ \bar\psi\end{matrix}\right)}{}
= -i\sigma_3 \frac{\partial}{\partial x}(2 {\cal M}_0)^{-1}
\big(E_0 + E_1 (2{\cal M}_0) +\dots + E_{N+1}(2{\cal M}_0)^{N+1}+\cdots \big) \left( \begin{matrix} \psi \\ \bar\psi \end{matrix}\right).
\end{gather*}
Shortly it is
\begin{gather*}
i\sigma_3 \frac{\partial}{\partial t}\left( \begin{matrix} \psi \\ \bar\psi\end{matrix}\right) = -i\sigma_3 \frac{\partial}{\partial x}
(2 {\cal M}_0)^{-1} H(2{\cal M}_0) \left( \begin{matrix} \psi \\ \bar\psi \end{matrix}\right).
\end{gather*}
By replacing ${\cal M}_0 \rightarrow {\cal M}$, f\/inally we get the DNLS nonlinearization of the linear Schr\"odinger equation with arbitrary form of dispersion (\ref{dispersion}),
\begin{gather}
i\sigma_3 \frac{\partial}{\partial t}\left( \begin{matrix} \psi \\ \bar\psi\end{matrix}\right) = -i\sigma_3 \frac{\partial}{\partial x}
(2 {\cal M})^{-1} H(2{\cal M}) \left( \begin{matrix} \psi \\ \bar\psi \end{matrix}\right).\label{NDNLS}
\end{gather}
We notice that another type of nonlinearization of the linear Schr\"odinger equation, based on the NLS hierarchy was derived in~\cite{Pashaev}.
\subsubsection{DNLS hierarchy linear problem}\label{section4.2.1}
For the $N$-th f\/low we start with
\begin{gather*}
\left( \begin{matrix} C_N \\ B_N\end{matrix}\right) = \left( \begin{matrix} \lambda^{N+1} & 0 \\ 0 & \lambda^N\end{matrix}\right)
[N+1]_{L/\lambda}\left( \begin{matrix} \kappa^2 \bar\psi \\ \psi\end{matrix}\right)\\
\hphantom{\left( \begin{matrix} C_N \\ B_N\end{matrix}\right)}{} = \left( \begin{matrix} 0 & \lambda^{N+1} \\ \lambda^N & 0\end{matrix}\right)
\sigma_1 \left(I + \frac{L}{\lambda} + \dots + \frac{L^N}{\lambda^N}\right)\sigma_1\left( \begin{matrix} \psi \\ \kappa^2 \bar\psi \end{matrix}\right)\\
\hphantom{\left( \begin{matrix} C_N \\ B_N\end{matrix}\right)}{}
= \left( \begin{matrix} 0 & \lambda^{N+1} \\ \lambda^N & 0\end{matrix}\right)
\left(I + \frac{{\cal L}}{\lambda} + \dots + \frac{{\cal L}^N}{\lambda^N}\right)\left( \begin{matrix} \psi \\ \kappa^2 \bar\psi \end{matrix}\right).
\end{gather*}
Replacing ${\cal L}$ by ${\cal M}$
\begin{gather*}
{\cal L } = \left( \begin{matrix} 1 & 0 \\ 0 & \kappa^2\end{matrix}\right){\cal M}\left( \begin{matrix} 1 & 0 \\ 0 & \kappa^2\end{matrix}\right)
\end{gather*}
f\/inally we get
\begin{gather*}
\left( \begin{matrix} C_N \\ B_N\end{matrix}\right) = \left( \begin{matrix} 0 & \kappa^2\lambda^{N+1} \\ \lambda^N & 0\end{matrix}\right)
[N+1]_{{\cal M}/\lambda}\left( \begin{matrix} \psi \\ \bar\psi\end{matrix}\right)
\end{gather*}
or
\begin{gather*}
\left( \begin{matrix} C_N \\ B_N\end{matrix}\right) = \left( \begin{matrix} 0 & \kappa^2\lambda \\ 1 & 0\end{matrix}\right)
\frac{{\cal M}^{N+1} - \lambda^{N+1}}{{\cal M} - \lambda}\left( \begin{matrix} \psi \\ \bar\psi\end{matrix}\right).
\end{gather*}
Similar calculations for $A$ give
\begin{gather*}
A_N = \lambda^{N+1} + i\kappa^2\big({-}\partial^{-1}\bar \psi, \partial^{-1}\psi\big) \lambda {\cal M} \frac{{\cal M}^{N} - \lambda^{N}}{{\cal M} - \lambda} \left( \begin{matrix} \psi \\ \bar\psi\end{matrix}\right).
\end{gather*}
Here we like to emphasize that these expressions are written in terms of two base operator $pq$-numbers
with basis ${\cal M}$ and $\lambda$
\begin{gather*}
[N]_{{\cal M}, \lambda} = \frac{{\cal M}^{N} - \lambda^{N}}{{\cal M} - \lambda}.
\end{gather*}
\subsubsection{Arbitrary dispersion linear problem}\label{section4.2.2}
To construct the linear problem for arbitrary dispersive DNLS (\ref{NDNLS}) with dispersion (\ref{dispersion}) we f\/irst expand
\begin{gather*}
i\sigma_3 \frac{\partial}{\partial t}\left( \begin{matrix} \psi \\ \bar\psi\end{matrix}\right)
= -i\sigma_3 \frac{\partial}{\partial x}
(2 {\cal M})^{-1} \big(E_0 + E_1(2{\cal M}) + \dots + E_N(2{\cal M})^N+\cdots \big) \left( \begin{matrix} \psi \\ \bar\psi \end{matrix}\right)\\
\hphantom{i\sigma_3 \frac{\partial}{\partial t}\left( \begin{matrix} \psi \\ \bar\psi\end{matrix}\right) }{}
= -i\sigma_3 \frac{\partial}{\partial x}
 \big(E_0 (2 {\cal M})^{-1} + E_1 + \dots + E_{N+1}(2{\cal M})^N+\cdots \big) \left( \begin{matrix} \psi \\ \bar\psi \end{matrix}\right)\\
\hphantom{i\sigma_3 \frac{\partial}{\partial t}\left( \begin{matrix} \psi \\ \bar\psi\end{matrix}\right) }{}
=\left(\frac{E_0}{2}\left(-i\sigma_3 \frac{\partial}{\partial x}\right){\cal M}^{-1} + E_1 \left(-i\sigma_3 \frac{\partial}{\partial x}\right) + \cdots\right.\\
\left.
\hphantom{i\sigma_3 \frac{\partial}{\partial t}\left( \begin{matrix} \psi \\ \bar\psi\end{matrix}\right)= }{}
+ 2^N E_{N+1}\left(-i\sigma_3 \frac{\partial}{\partial x}\right)({\cal M})^N+\cdots \right) \left( \begin{matrix} \psi \\ \bar\psi \end{matrix}\right)\\
\hphantom{i\sigma_3 \frac{\partial}{\partial t}\left( \begin{matrix} \psi \\ \bar\psi\end{matrix}\right) }{}
= \left(\frac{E_0}{2}i\sigma_3 \frac{\partial}{\partial t_{-1}} + E_1 i\sigma_3 \frac{\partial}{\partial t_0} + \dots + 2^N E_{N+1}i\sigma_3 \frac{\partial}{\partial t_N} +\cdots \right) \left( \begin{matrix} \psi \\ \bar\psi \end{matrix}\right).
\end{gather*}
Then we def\/ine new time variable
\begin{gather*}
\frac{\partial}{\partial t}= \frac{E_0}{2} \frac{\partial}{\partial t_{-1}} + E_1 \frac{\partial}{\partial t_0} + \dots + 2^N E_{N+1} \frac{\partial}{\partial t_N} +\dots = \sum^\infty_{N=0} E_N 2^{N-1}\frac{\partial}{\partial t_{N-1}}
\end{gather*}
and the linear problem with
\begin{gather*}
U = \left( \begin{matrix} \lambda & \psi \\ \lambda \kappa^2 \bar\psi & -\lambda \end{matrix} \right),\qquad
V = \left( \begin{matrix} A & B \\ C & -A \end{matrix} \right),
\end{gather*}
where
\begin{gather*}
V = \sum^\infty_{N=0} E_N 2^{N-1} V_{N-1}.
\end{gather*}
Substituting expressions obtained in previous section we f\/ind
\begin{gather*}
\left( \begin{matrix} C \\ B\end{matrix}\right) = \left( \begin{matrix} 0 & \kappa^2 \lambda \\ 1 & 0\end{matrix}\right)
\frac{H(2{\cal M}) - H(2\lambda)}{2{\cal M} - 2\lambda}\left( \begin{matrix} \psi \\ \bar\psi\end{matrix}\right),\\
A = \frac{1}{2} H(2\lambda) + i\kappa^2 \big({-}\partial^{-1}\bar\psi, \partial^{-1}\psi\big)\frac{\lambda H(2{\cal M}) - {\cal M}H(2\lambda)}{2{\cal M} - 2\lambda}\left( \begin{matrix} \psi \\ \bar\psi\end{matrix}\right).
\end{gather*}
\section{Relativistic DNLS}\label{section5}
As an application of the above procedure, in this section we consider the relativistic DNLS equation determined by semirelativistic
dispersion
\begin{gather*}
H(p) = \sqrt{m^2 c^4 + p^2 c^2} = mc^2 \sqrt{1 + \frac{p^2}{m^2 c^2}}.
\end{gather*}
Following to this procedure we f\/ind then RDNLS as
\begin{gather*}
i\sigma_3 \frac{\partial}{\partial t}\left( \begin{matrix} \psi \\ \bar\psi\end{matrix}\right) = -i\sigma_3 \frac{\partial}{\partial x}
\frac{mc^2}{2} {\cal M}^{-1} \sqrt{1 + \frac{4}{m^2 c^2}{\cal M}^2} \left( \begin{matrix} \psi \\ \bar\psi \end{matrix}\right),
\end{gather*}
and the corresponding linear problem
\begin{gather*}
\left( \begin{matrix} C \\ B\end{matrix}\right) = \left( \begin{matrix} 0 & \kappa^2 \lambda \\ 1 & 0\end{matrix}\right)
mc^2 \frac{\sqrt{1 + \frac{4}{m^2 c^2}{\cal M}^2} - \sqrt{1 + \frac{4}{m^2 c^2}{\lambda}^2} }{2{\cal M} - 2\lambda}\left( \begin{matrix} \psi \\ \bar\psi\end{matrix}\right),\\
A = \frac{1}{2}mc^2 \sqrt{1 + \frac{4}{m^2 c^2}{\lambda}^2} \\
\hphantom{A =}{}
+ i\kappa^2 \big({-}\partial^{-1}\bar\psi, \partial^{-1}\psi\big) mc^2 \frac{\lambda \sqrt{1 + \frac{4}{m^2 c^2}{\cal M}^2} - {\cal M}\sqrt{1 + \frac{4}{m^2 c^2}{\lambda}^2} }{2{\cal M} - 2\lambda} \left( \begin{matrix} \psi \\ \bar\psi\end{matrix}\right).
\end{gather*}
Here the operator
\begin{gather*}
{\cal M}^{-1} = 2\left( \begin{matrix} i\partial^{-1} - \kappa^2\partial^{-1} \psi \partial^{-1} \bar\psi& \kappa^2\partial^{-1} \psi \partial^{-1} \psi \\ \kappa^2\partial^{-1} \bar\psi \partial^{-1} \bar\psi &
-i\partial^{-1} - \kappa^2\partial^{-1} \bar\psi \partial^{-1} \psi\end{matrix}\right)
\end{gather*}
and
\begin{gather*}
{\cal M}^{-1} \left( \begin{matrix} \psi \\ \bar\psi\end{matrix}\right) = 2i \sigma_3 \partial^{-1}\left( \begin{matrix} \psi \\ \bar\psi\end{matrix}\right).
\end{gather*}
Using the last expression and the formal expansion of the square root, we have the series of relativistic corrections to the nonrelativistic DNLS
\begin{gather*}
i\sigma_3 \frac{\partial}{\partial t}\left( \begin{matrix} \psi \\ \bar\psi\end{matrix}\right) =
\left(mc^2 -i\sigma_3 \frac{\partial}{\partial x}\left(\frac{{\cal M}}{m} - \frac{{\cal M}^3}{m^3 c^2} + 2 \frac{{\cal M}^5}{m^5 c^4} +\cdots \right)\right)\left( \begin{matrix} \psi \\ \bar\psi\end{matrix}\right).
\end{gather*}
For $m=1$ it gives the f\/irst few terms of the relativistic corrections as
\begin{gather*}
i\sigma_3 \frac{\partial}{\partial t}\left( \begin{matrix} \psi \\ \bar\psi\end{matrix}\right) = c^2 \left( \begin{matrix} \psi \\ \bar\psi\end{matrix}\right) -\frac{1}{2}\frac{\partial^2}{\partial x^2}\left( \begin{matrix} \psi \\ \bar\psi\end{matrix}\right) + \frac{\kappa^2}{2} i\sigma_3\frac{\partial}{\partial x} |\psi|^2\left( \begin{matrix} \psi \\ \bar\psi\end{matrix}\right)\\
\hphantom{i\sigma_3 \frac{\partial}{\partial t}\left( \begin{matrix} \psi \\ \bar\psi\end{matrix}\right) =}{} -i\sigma_3 \frac{\partial}{\partial x} \left(- \frac{{\cal M}^3}{c^2} + 2 \frac{{\cal M}^5}{c^4} +\cdots \right) \left( \begin{matrix} \psi \\ \bar\psi\end{matrix}\right).
\end{gather*}
The whole set of relativistic corrections is given by the following inf\/inite sum
\begin{gather*}
-i\sigma_3 \frac{\partial}{\partial x}\sum^\infty_{n=2}(-1)^{n+1} \frac{(2n)!}{4^n (n!)^2(2n-1)}\frac{(2 {\cal M})^{2n-1}}{c^{2n-2}} \left( \begin{matrix} \psi \\ \bar\psi\end{matrix}\right) .
\end{gather*}

The relativistic corrections at any order in this sum represent an integrable system and allow us to speculate on possible physical applications of our relativistic DNLS equation. One possible application is related with long-wavelength dynamics of dispersive Alfven waves, propagating along an ambient magnetic f\/ield in relativistic plasma. So that for every relativistic correction to the wave dispersion we can develop an integrable model at any order in~$1/c^2$. As a~next possible application we can mention the problem of one-dimensional relativistic anyons and the relativistic chiral solitons propagation in quantum potential, where possible to have an accurate description of relativistic corrections to dispersion and nonlinearity, preserving integrability. One more application could be related with relativistic corrections to the modif\/ied KP equation and corresponding resonant solitons as some type of relativistic modif\/ied KP equation from modif\/ied KP hierarchy.

\section{Conlusions}\label{section6}

In the present paper we have derived Lax representation for the ${\rm SL}(2,{\mathbb R})$ KN hierarchy and DNLS hierarchy by using operator $q$-numbers and $q$-derivatives with two bases, one of which is the recursion operator and the another one is the spectral parameter. These compact expressions allowed us to derive DNLS with an arbitrary dispersion. Choosing the semirelativistic form of the dispersion we have constructed relativistic DNLS, which becomes DNLS in nonrelativistic limit and it is integrable at any order of relativistic corrections to DNLS. Since the $U$ operator for this equation is in the same form as for DNLS, the spectral characteristics for both models without time evolution would be the same. But in time evolution of solitons we will have now the relativistic form of dispersion.

\subsection*{Acknowledgements}
This work was partially supported by Izmir Institute of Technology, Turkey, and Institute of Mathematics, Academia Sinica, Taiwan. The work of O.K.P.\ was supported by the Scientif\/ic and Technological Research Council of Turkey (T\"UBITAK), Grant No: TBAG-116F206.

\pdfbookmark[1]{References}{ref}
\LastPageEnding

\end{document}